\pdfoutput=1

\documentclass[conference,compsoc]{IEEEtran}
\usepackage{times}           
\usepackage[T1]{fontenc}
\usepackage[utf8]{inputenc}
\usepackage{longtable}
\usepackage[table]{xcolor}
\usepackage{adjustbox}

\usepackage{booktabs}        
\usepackage{tabularx}        
\usepackage{multirow}        
\usepackage{array}           
\usepackage{ragged2e}        
\usepackage{colortbl}        
\usepackage{xcolor}          
\usepackage{makecell}        
\usepackage{pdflscape}       
\usepackage{amsmath}   
\usepackage{amsfonts}  
\usepackage{rotating}
\usepackage{cuted}
\usepackage{capt-of}

\usepackage{setspace}
\usepackage{caption}
\usepackage{url}
\usepackage{xurl}
\usepackage{tikz}
\usetikzlibrary{arrows.meta,positioning,calc,shapes.geometric}
\usetikzlibrary{positioning,arrows.meta,calc}
\usepackage{stfloats}

\newcolumntype{L}[1]{>{\raggedright\arraybackslash}p{#1}}
\newcolumntype{Y}{>{\raggedright\arraybackslash}X}

 \usepackage{hyperref}
 
\newcolumntype{P}[1]{>{\RaggedRight\arraybackslash\small}p{#1}}
 
\usepackage{longtable}   
\usepackage{array}       
\usepackage{booktabs}
\usepackage{tabularx}
\usepackage{amssymb}
\usepackage{enumitem}
\usepackage{multirow}

\newtheorem{definition}{Definition}[section]

\definecolor{p1col}{RGB}{31,78,121}    
\definecolor{p3col}{RGB}{31,78,121}    
\definecolor{rootcol}{RGB}{166,28,49}  
\definecolor{seccol}{RGB}{120,120,120} 
\definecolor{ampcol}{RGB}{200,200,200} 
\definecolor{domfill}{RGB}{235,240,247}

\usetikzlibrary{arrows.meta, positioning, calc}
\tikzset{
  dom/.style={draw=p1col, fill=domfill, rounded corners=1.5pt, minimum height=5.5mm,
              minimum width=12.5mm, font=\scriptsize, inner sep=1.5pt},
  domatk/.style={dom, draw=rootcol, fill=rootcol!8},
  tbline/.style={dashed, gray!70, line width=0.5pt},
  pathArrow/.style={-{Stealth[length=2.2mm]}, line width=0.9pt, rootcol},
  p1mark/.style={circle, fill=p1col, draw=p1col, inner sep=0pt, minimum size=2.6mm},
  p1root/.style={circle, fill=rootcol, draw=rootcol, inner sep=0pt, minimum size=2.6mm},
  p3mark/.style={circle, fill=white, draw=p3col, line width=0.9pt, inner sep=0pt, minimum size=2.6mm},
  p2mark/.style={diamond, fill=seccol!35, draw=seccol, inner sep=0pt, minimum size=2.4mm},
  propArrow/.style={-{Stealth[length=2.4mm]}, line width=1.1pt, p1col},
  carryArrow/.style={-{Stealth[length=2mm]}, line width=0.8pt, rootcol, dashed},
  lanelabel/.style={font=\scriptsize\bfseries, anchor=east},
  note/.style={font=\tiny, text=black!75},
  valbox/.style={draw=gray!60, fill=gray!6, rounded corners=1pt, font=\tiny,
                 inner sep=2pt, align=left},
}
\AtBeginDocument{%
  }

\begin{document}
%

\title{Trust Boundary Semantic Gaps: A Multi-dimensional Analysis and Mitigation for Security-by-Design} 

\author{\IEEEauthorblockN{Doyeon Kim, Jin-Young Choi, and Junghee Lee\textsuperscript{*}}
\IEEEauthorblockA{Korea University\\
Seoul, Republic of Korea\\
\textsuperscript{*}Corresponding author}}

\maketitle

\begin{abstract}
    Modern systems use format-, protocol-, and signature-based mechanisms before accepting artifacts across trust boundaries. These mechanisms are necessary: they show that an artifact is well formed, protocol-compliant, or properly authenticated. They do not, however, show that the artifact satisfies the semantic security properties required by the receiving domain. A signed update or an authenticated token may therefore be accepted yet enable compromise.
We call this condition a Trust Boundary Semantic Gap (TBSG): an artifact crosses a trust boundary and passes correctly implemented syntactic validation, but the assertions established by that pass are insufficient to satisfy the receiving domain's security requirements. TBSG concerns what remains unestablished after a syntactic pass, not absent checks or implementation bugs.
Analyzing 75 publicly reported security incidents (2014--2025) at the boundary level, we organize semantic misalignment into a four-dimensional analysis model: Identity, Spatial, Temporal, and Interpretation (MDTBSG). Building on it, we develop Trust Boundary Semantic Analysis and Mitigation (TBSAM), a design-time framework that identifies TBSGs from design specifications, prioritizes them, traces propagated gaps to their originating boundary, and maps each to candidate architectural controls.
We apply TBSAM to a retrospective reconstruction of the SolarWinds/SUNBURST supply-chain attack, showing how it makes receiving-domain assumptions explicit, separates locally originating from propagated gaps, and identifies controls that interrupt the path. These results suggest that syntactic validation, while necessary, is not sufficient at trust boundaries, and that making trust-boundary assumptions explicit can complement Security-by-Design.

\end{abstract}


%
\IEEEpeerreviewmaketitle

\section{Introduction}
\label{sec:introduction}

Modern systems rarely accept artifacts without validation. When a message,
token, software package, API request, or update crosses a trust boundary, the
receiving domain commonly applies syntactic validation: signatures and hashes
for software updates, schema and protocol checks for API requests, and tokens
or credentials for sessions. These checks are necessary. They reject malformed
artifacts, enforce protocol rules, and establish limited assertions such as
integrity under an accepted key, format conformance, or credential validity.
They do not, however, establish every security property that the receiving
domain may rely on when it processes the artifact.

Major incidents show this gap. In SolarWinds/SUNBURST, a valid SolarWinds
signature did not establish that the build process produced the binary from
the intended source. In Log4Shell, syntactically valid input did not establish
that logged data would remain inert. In the Capital One breach,
protocol-compliant requests did not establish that the request should reach
the cloud metadata service or that the resulting authority should be usable.
These incidents differ in mechanism, but share the same boundary-level
structure: an artifact passed syntactic validation at a trust boundary, and
the receiving domain processed it under a security assumption that the
validation step had not established.

We define this condition as a \emph{Trust Boundary Semantic Gap} (TBSG): an
artifact crosses a trust boundary and passes correctly implemented syntactic
validation, but the assertions established by that pass do not satisfy the
security properties the receiving domain requires. A TBSG is not an absent
validation step, a parser bug, or a failed signature check. It captures a semantic gap between the assertions established by syntactic validation and the security meaning that the receiving domain later relies on.

Existing approaches explain important parts of this setting. STRIDE-based
threat modeling helps security architects identify threat categories and map
them to controls. Input-validation guidance explains how to reject malformed
inputs. Prior work on semantic gaps studies meaning mismatches inside specific
settings. This paper asks a residual boundary question: after a correctly
implemented validation step succeeds at a trust boundary, which
receiving-domain security property remains unestablished?

To answer this question, we analyze 75 publicly reported security incidents
from 2014 to 2025 at the level of trust-boundary crossings. For each boundary,
we record the artifact, the syntactic validation applied by the receiving
domain, the assertion established by the pass, and the property left open.
From this analysis, we organize recurring forms of semantic misalignment into
\emph{MDTBSG}, a four-dimensional analysis model: Identity (\emph{who} is
behind an artifact), Spatial (\emph{where} it can reach), Temporal
(\emph{when} an assertion remains valid), and Interpretation (\emph{what}
behavior it causes).

Identifying a gap at a trust boundary is only the first step. Security
architects still need a procedure that turns the gap into a design decision.
We therefore develop \emph{Trust Boundary Semantic Analysis and Mitigation}
(TBSAM), a design-time framework composed of a four-stage procedure. TBSAM
builds a boundary record from design specifications, selects candidate
dimensions, prioritizes gaps, maps them to architectural controls, and traces
propagated gaps back to the preceding boundary where they first appear.

TBSAM is not a runtime detector and does not replace threat modeling. It
starts from the controls that threat modeling selects and examines what those
controls leave open. We apply TBSAM to a post-incident reconstruction of
SolarWinds/SUNBURST. The analysis identifies three locally originating gaps
and one propagated gap. A control placed where the key gap first
appears---source-to-binary attestation at the build-output boundary---breaks
the reconstructed path before signing. We do not claim that TBSAM
automatically infers security properties; rather, it structures a security
architect's review of what remains unestablished after a selected boundary
control succeeds.

\smallskip
Our contributions are summarized as follows.
\begin{itemize}
\item A design-level condition where syntactic validation succeeds at a trust boundary, but the established assertions do not satisfy the security properties required by the receiving domain.

\item A four-dimensional analysis model of semantic misalignment, built from 75 publicly reported security incidents and organized around Identity, Spatial, Temporal, and Interpretation.

\item A design-time framework composed of a four-stage procedure that identifies TBSGs from design specifications, prioritizes them, and maps them to architectural controls, with a SolarWinds/SUNBURST case study and STRIDE-based comparison over the same input.
\end{itemize}

\section{Background and Motivation}
\label{sec:related}

\subsection{Trust Boundary in Software Security}

A trust boundary is commonly understood as a boundary between components
or domains that operate under different security policies, privileges, or
trust levels. Microsoft STRIDE and OWASP Threat Modeling use data flow
diagrams (DFDs) to reason about such boundaries. In DFDs, trust boundaries
are represented as lines or boundary markers that indicate a change in trust
as data flows through the system~\cite{owasp_threat_modeling}. In this
practice, the analyst examines each DFD element or boundary crossing,
enumerates the relevant threat categories, and maps each finding to a
conventional control, such as authentication for Spoofing or code signing
for Tampering~\cite{kohnfelder1999threats,shostack2014threat}.

The same per-element practice has been extended to the software supply
chain, where the SLSA framework defines integrity levels for build and
distribution pipelines~\cite{slsa}. Prior work has also studied trust
boundaries in concrete settings, including network-level
boundaries~\cite{rose2020nist}, process isolation~\cite{sammler2019high},
and API gateways~\cite{nist_sp800_228}.

In this paper, we treat a trust boundary as a security transition between
two trust domains that operate under different security policies. The
crossing may occur through a concrete interface-level event, such as an API
call, IPC message, signed update, token, or request. The boundary itself,
however, is not the interface but the change in security assumptions between
the sending and receiving domains. When an artifact crosses such a boundary,
the receiving domain must determine what the artifact actually proves before
relying on it.

\subsection{Syntactic Validation}

Syntactic validation checks whether data satisfies expected structural
rules, follows a required format, or conforms to a protocol or data standard
~\cite{ssac_sac058}. It differs from semantic validation, which checks
whether a value is correct in a specific context. OWASP makes a similar
distinction: syntactic validation checks the syntax of structured fields,
while semantic validation checks whether their values are correct in a
specific business context~\cite{owasp_input_validation}.

In this paper, we use \emph{syntactic validation} to refer to checks that
establish structural or representational properties of an artifact at a trust
boundary. Prior work uses the term in this sense. Acto treats schema, type,
range, and pattern checks as syntactic constraints~\cite{acto}; Crackers
uses syntactic checks as structural pre-filtering before semantic
analysis~\cite{crackers}; and Nail shows that mixing syntactic validation
with semantic processing can introduce vulnerabilities~\cite{nail}.

At trust boundaries, two additional classes of checks often play the same
limited role, even when prior work does not always call them syntactic
validation. We treat them as syntactic in scope because they establish
properties about the artifact's representation, integrity, or protocol form,
but do not establish whether the artifact satisfies the semantic security
properties required by the receiving domain.

\textit{Cryptographic integrity checks}---such as signature verification,
HMAC verification, or certificate-chain checking---establish that an artifact
has not been modified and is bound to a known key. They do not, by themselves,
establish that the signer is trustworthy in the receiving domain or that the
signed content is safe to act on.

\textit{Protocol compliance checks}---such as HTTP header checking, URI
syntax checking, or API parameter type checking---establish that a message
follows expected protocol rules~\cite{owasp_input_validation}. They leave open
whether the request is directed to an intended destination or whether the
requested operation is permitted in the receiving domain.

The common limitation is that syntactic validation establishes only a limited
assertion: the artifact satisfies the checks that the boundary performs. NIST
SP 800-228 makes this point directly for input handling: even input that
satisfies syntactic validation may still be malicious if it attempts to make
the system misbehave~\cite{nist_sp800_228}.

\subsection{Semantic Gaps and Semantic Attacks in Security Analysis}

In virtual machine introspection, an external monitor observes low-level
machine state while security policies are expressed in higher-level OS
concepts. Bridging this mismatch---interpreting raw bytes as guest semantic
state---is the classic semantic-gap problem
~\cite{garfinkel2003vmi,fu2012space,jain2014sok}. A similar mismatch appears in web application security. Su and Wassermann
show that command injection attacks exploit a gap between how an application
treats input as character data and how a downstream interpreter treats the
same input as commands~\cite{su2006essence}. Network protocol research has
also identified semantic gaps in cross-layer data processing, where different
protocol layers interpret the same data under different assumptions or fail
to handle exceptional cases consistently~\cite{10.1145/3689819}.

Across these settings, the details differ, but the common structure is the
same: one side assigns or observes a meaning that does not match the meaning
required or intended by another side. Prior work studies this structure within
specific settings, such as VMI, web applications, and network protocols.

The term semantic attack is used in a related but broader line of work.
Schneier describes semantic attacks as attacks that target the meaning and
interpretation of information, rather than only its physical representation
or syntactic structure. He also notes that cryptographic mechanisms such as
signatures, authentication, and integrity checks do not by themselves solve
this class of attack, because they can protect the form or origin of
information without guaranteeing that the interpreted meaning is safe
~\cite{10.1145/355112.355131}. Heartfield and Loukas study semantic attacks in user-facing systems, including phishing, file masquerading, and spoofed websites, where attackers
manipulate the user-computer interface to compromise information security
~\cite{10.1145/2835375}. Defense work also reflects this concern: Razzaq et
al. argue that signature-based detection can miss attacks whose meaning
depends on context, and propose semantic rules and ontologies to describe the
context of web attacks~\cite{10.1016/j.ins.2013.08.007}.

\subsection{Motivation: Syntactic Pass, Semantic Failure}

The preceding subsections reviewed three established lines of work that this
paper builds on. Trust-boundary work explains where security boundaries are
placed and what checks should be applied when data, calls, or artifacts cross
them. Work on syntactic validation explains how a system checks structural or
representational properties of a crossing artifact, while also showing that
the assertions established by such checks are limited. Work on semantic gaps
and semantic attacks shows that meaning-level mismatches can lead to security
failures even when lower-level checks appear to succeed.

Together, these lines of work leave a boundary-level failure case
insufficiently explained. Semantic-gap work usually defines and addresses the
gap within a specific setting, such as VMI, web applications, or network
protocols. Trust-boundary work mainly focuses on where boundaries are placed
and what checks should be deployed at them. Work on syntactic validation often
focuses on local input handling. We are not aware of prior work that directly
treats the following case as a general design-level condition: syntactic
validation works correctly at a trust boundary, the artifact passes the
boundary check, but the receiving domain still relies on a semantic security
property that the syntactic pass does not establish.

This gap is not only conceptual: the incidents sketched in
Section~\ref{sec:introduction}---SolarWinds, Log4Shell, and Capital
One---each pass the deployed syntactic gate while exposing the receiving
domain to an unsafe artifact.

The next section defines this condition as a Trust Boundary Semantic Gap
(TBSG). We then study a set of 75 real-world security incidents to show that
this is not an isolated accident, but a recurring structural pattern across
different systems.

\section{Trust Boundary Semantic Gap}
\label{sec:tbsg}

This section develops TBSG from observed security failures in four steps. Section~\ref{subsec:tbsg_pattern} describes the recurring pattern: an artifact passes syntactic validation at a trust boundary but fails to satisfy the security properties required by the receiving domain. Section~\ref{subsec:incident_analysis} analyzes 75 real-world incidents at the trust boundary level. Section~\ref{subsec:tbs_definition} formalizes trust boundary and TBSG. Section~\ref{subsec:MDTBSG} organizes recurring forms of semantic misalignment into four dimensions: Identity, Spatial, Temporal, and Interpretation.

\subsection{From Syntactic Pass to Semantic Failure}
\label{subsec:tbsg_pattern}
Modern software systems consist of components that operate under different security assumptions. We refer to a set of components sharing a common security assumption as a trust domain, and to the transition an artifact makes between two domains as a trust boundary. At a boundary, the receiving domain typically applies syntactic validation—structural, format, integrity, protocol, or delivery checks—before accepting the artifact. These checks are necessary, but they do not by themselves establish the receiving domain's security requirements.

The incidents analyzed in Section~\ref{subsec:incident_analysis} show a recurring pattern: an artifact passes the boundary's syntactic validation, yet the receiving domain still lacks assurance about its origin, scope, freshness, or meaning. The failure does not stem from absent validation or an implementation flaw. We call this condition a Trust Boundary Semantic Gap (TBSG): the assertions established by a syntactic pass are insufficient to satisfy the receiving domain's security requirements.

%

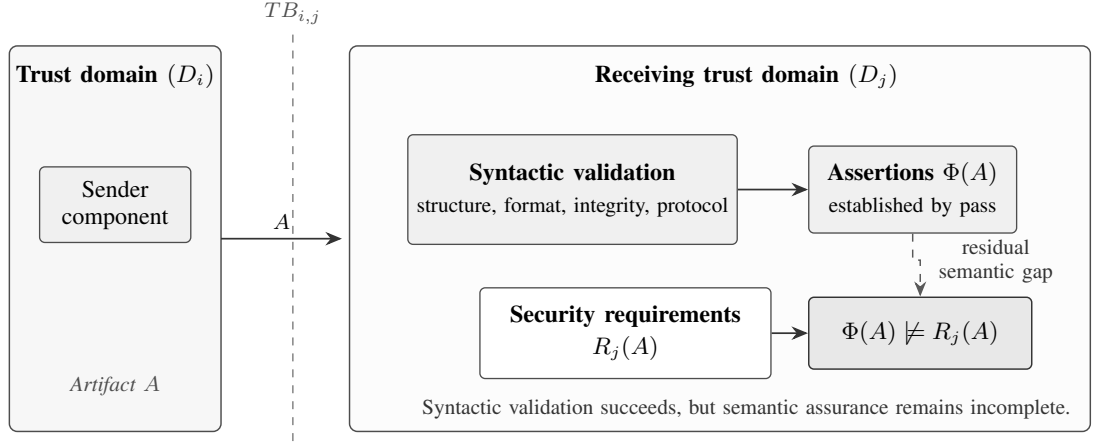
\begin{figure*}[t]
\centering
\begin{tikzpicture}[
  x=1cm, y=1cm,
  font=\small,
  domnode/.style={
    draw=black!70,
    rounded corners=3pt,
    line width=0.55pt,
    fill=black!3,
    minimum width=2.8cm,
    minimum height=5.1cm
  },
  procbox/.style={
    draw=black!70,
    rounded corners=2pt,
    line width=0.5pt,
    fill=black!6,
    align=center
  },
  reqbox/.style={
    draw=black!70,
    rounded corners=2pt,
    line width=0.5pt,
    fill=white,
    align=center
  },
  emphbox/.style={
    draw=black!80,
    rounded corners=2pt,
    line width=0.55pt,
    fill=black!9,
    align=center
  },
  arr/.style={
    -{Stealth[length=2.2mm,width=1.8mm]},
    line width=0.65pt,
    draw=black!80
  },
  gaparr/.style={
    -{Stealth[length=2mm,width=1.6mm]},
    dashed,
    line width=0.55pt,
    draw=black!65
  },
  tbline/.style={
    dashed,
    line width=0.6pt,
    draw=black!45
  },
  title/.style={
    font=\small\bfseries,
    align=center
  },
  note/.style={
    font=\footnotesize,
    align=center,
    text=black!80
  },
  lbl/.style={
    font=\footnotesize\itshape,
    text=black!70
  }
]

\node[domnode] (Di) at (0,0) {};
\node[title] at (0,2.15)
  {Trust domain $(D_i)$};

\node[procbox, minimum width=2.0cm, minimum height=1.0cm]
  (sender) at (0,0.45)
  {Sender\\component};

\node[lbl] at (0,-1.95)
  {Artifact $A$};

\draw[tbline] (2.35,2.70) -- (2.35,-2.70);

\node[lbl, anchor=south] at (2.35,2.72)
  {$TB_{i,j}$};

\draw[arr] (1.40,0) -- node[above, font=\footnotesize] {$A$} (3.00,0);

\coordinate (recvNW) at (3.10, 2.55);
\coordinate (recvSE) at (12.90,-2.55);

\draw[
  draw=black!70,
  rounded corners=3pt,
  line width=0.55pt,
  fill=black!1
] (recvNW) rectangle (recvSE);


\node[title] at (8.35,2.15)
  {Receiving trust domain $(D_j)$};

\node[procbox, minimum width=3.25cm, minimum height=1.45cm]
  (syn) at (6.05,0.65)
  {\textbf{Syntactic validation}\\[2pt]
   \footnotesize structure, format, integrity, protocol};

\node[procbox, minimum width=2.75cm, minimum height=1.15cm]
  (phi) at (10.55,0.65)
  {\textbf{Assertions} $\Phi(A)$\\[2pt]
   \footnotesize established by pass};

\draw[arr] (syn.east) -- (phi.west);

\node[reqbox, minimum width=3.85cm, minimum height=1.20cm]
  (req) at (6.75,-1.25)
  {\textbf{Security requirements}\\[2pt]
   $R_j(A)$};

\node[emphbox, minimum width=2.95cm, minimum height=0.95cm]
  (gap) at (10.65,-1.25)
  {$\Phi(A) \not\models R_j(A)$};

\draw[arr] (req.east) -- (gap.west);

\draw[gaparr] (phi.south) -- ++(0,-0.35) -| (gap.north);

\node[note] at (11.65,-0.3)
  {residual\\semantic gap};

\node[
  note,
  text width=8.7cm,
  align=center
] at (8.35,-2.25)
  {Syntactic validation succeeds, but semantic assurance remains incomplete.};

\end{tikzpicture}

\caption{Structural model of a Trust Boundary Semantic Gap (TBSG). An artifact
$A$ crosses trust boundary $TB_{i,j}$ and passes syntactic validation, which
establishes assertions $\Phi(A)$. A TBSG arises when these assertions are
insufficient to satisfy the receiving domain's security requirements $R_j(A)$,
that is, when $\Phi(A) \not\models R_j(A)$.}
\label{fig:tbsg-model}
\end{figure*}

Figure~\ref{fig:tbsg-model} illustrates this structure. A sending domain passes an artifact to a receiving domain. The receiving domain applies syntactic validation, and the artifact passes. The pass establishes only the properties covered by the boundary check. The receiving domain may still need additional security assurance before it can process the artifact safely. If the design does not identify this gap, the receiving domain may treat the artifact as safe even though its security requirements remain incomplete.

\subsection{Incident Set and Boundary-level Analysis}
\label{subsec:incident_analysis}
This section examines how the TBSG pattern introduced in Section~\ref{subsec:tbsg_pattern} appears in real-world security incidents. We considered security incidents from 2014 to 2025 for which public sources provide evidence of real exploitation. These sources include the CISA Known Exploited Vulnerabilities catalog, CISA advisories, MITRE ATT\&CK entries, and technical reports from security vendors such as Mandiant and CrowdStrike.

We reviewed 81 candidate incidents and excluded six because the exploited trust boundary did not include syntactic validation as part of the design. These cases involved memory-safety violations, injection caused by missing sanitization, or implementation flaws in cryptographic primitives. They fall outside the scope of TBSG because the first validation gate was absent or broken. The remaining 75 incidents form the incident set analyzed in this section.

The incident set spans multiple platforms and attack classes. It includes operating systems such as Android, macOS, and Windows; cloud services such as AWS, Azure, and GCP; enterprise software such as Confluence, TeamCity, and Jenkins; network appliances such as Ivanti, FortiOS, and Cisco IOS XE; and software supply chains such as SolarWinds, XZ Utils, and 3CX. The attack vectors include supply-chain compromise, credential abuse, web application remote code execution, IPC-based privilege escalation, cloud identity misuse, and server-side request forgery.

We analyze each incident at the trust-boundary level rather than treating the incident as a single unit. For each incident, we trace the exploit path and record each trust boundary crossed by the artifact. For each boundary, we record three items: the crossing artifact, the syntactic validation applied by the receiving domain and the assertions it establishes, and the security property that remains unsatisfied after the artifact passes syntactic validation. For each incident, we assigned a dimension label only when the public record supported three facts at the relevant boundary: (1) the artifact crossed a trust boundary, (2) the receiving domain accepted the artifact after syntactic validation passed, and (3) the exploit depended on a receiving-domain security property that the syntactic pass did not establish. If the available public sources did not provide sufficient support for all three facts for a given dimension at a given boundary, that dimension was left unmarked. This rule is conservative: when a public report did not establish the boundary artifact, the syntactic pass, and the missing receiving-domain property, we did not assign the dimension. It applies to both positive labels and empty cells in Table~\ref{tab:Analysis}. Empty cells therefore indicate boundaries or dimensions for which the three-fact criterion was not met in the public record; they are part of the analysis, not an absence of analysis.

Across the 75 incidents, the same high-level structure appears repeatedly: an artifact crosses a trust boundary and passes syntactic validation, but the receiving domain still lacks a security property it needs to remain secure. This structure appears across different platforms, vendors, and attack vectors. It does not point to a single vulnerability class. Instead, it shows a recurring form of semantic misalignment at trust boundaries.

Table~\ref{tab:Analysis} summarizes the dimension labels assigned to each incident. Identity appears in 64 of 75 incidents (85.3\%) and Spatial in 62 of 75 (82.7\%); Temporal and Interpretation appear in 40 (53.3\%) and 59 (78.7\%), respectively. The prevalence of Identity and Spatial reflects a common structure of trust-boundary crossings: many artifacts are presented by, or on behalf of, a principal and are routed toward a resource or scope controlled by the receiving domain.

The eleven incidents without an Identity label are also informative. They occur at boundaries where the exploited step does not depend on a principal assertion, such as pure TOCTOU races, unauthenticated deserialization endpoints, pre-authentication memory corruption, or display-layer spoofing. In these cases, the primary gap lies in the Spatial, Temporal, or Interpretation dimension. This shows that MDTBSG does not assign dimensions by default. A dimension is marked only when the exploit mechanism depends on the property that the dimension captures and that property remains unestablished at the boundary.

Under this criterion, 5 incidents implicate a single dimension and 14 implicate two. Reinterpretation-only failures such as Trojan Source carry no Identity or Spatial label, while single-host code-execution incidents such as Spring4Shell or in-process deserialization (XStream) carry no Spatial label because the artifact's reach does not exceed the receiving host.

Before organizing these recurring forms, we first formalize the two concepts used throughout the analysis: trust boundary and TBSG (Section~\ref{subsec:tbs_definition}). Section~\ref{subsec:MDTBSG} then classifies the recurring forms into four dimensions.

\subsection{Trust Boundary and TBSG: Notation and Definitions}
\label{subsec:tbs_definition}
We introduce notation for the boundary-level analysis used in this paper. The notation makes the analysis explicit across system designs, but it does not define an automated decision procedure.

A software system consists of components that operate under different security policies and assumptions. We write a trust domain as $D_i$: a set of components that the system design treats under a common security policy and a common set of security assumptions. The set of trust domains is denoted by
\[
  \mathcal{D} = \{D_1, D_2, \ldots, D_n\}.
\]
A domain may be identified from the policy it enforces, the principals it trusts, the privileges it holds, the resources it can access, or the assumptions it makes about incoming artifacts. This abstraction does not require the implementation to explicitly encode domains.

For each domain $D_j$, we write $R_j$ for the security requirements that must hold for the domain to process artifacts safely. A security policy describes the rules that a domain applies; a security requirement describes what those rules are intended to protect. Examples include the authority of a principal, the allowed origin of a request, the valid lifetime of an assertion, and the intended meaning of an input.

We use the term \emph{artifact} $A$ to denote an input, message, token, software package, or response that moves from one domain to another. Examples include HTTP requests, authentication tokens, signed software packages, API responses, IPC messages, DNS responses, and cloud metadata responses. An artifact may carry assertions across domains, such as origin, identity, authority, freshness, format, or intended meaning. If $A$ passes, the validation establishes a set of assertions about $A$, denoted $\Phi(A)$. Because the enforced specification differs across boundaries, $\Phi(A)$ is relative to the boundary under analysis.

\begin{definition}[Trust Boundary]
\label{def:trust_boundary}
A trust boundary $TB_{i,j}$ exists when the system design permits at least one artifact, $A$, to cross from domain $D_i$ to domain $D_j$, and the two domains are distinct:
\[
  TB_{i,j} \iff
  \big(\exists A : D_i \xrightarrow{\,A\,} D_j\big)
  \wedge (i \neq j).
\]
\end{definition}

A trust boundary is directional. We analyze $TB_{i,j}$ and $TB_{j,i}$ separately because each direction may carry different artifacts, apply different syntactic validation, and impose different security requirements on the receiving domain. A policy difference alone does not create an active trust boundary; at least one artifact must cross between the domains.

When syntactic validation is present at $TB_{i,j}$, the receiving domain checks $A$ against the syntactic specification that the boundary is designed to enforce. We leave this specification implicit in the notation. If $A$ passes, syntactic validation establishes a set of assertions about $A$, denoted $\Phi(A)$.

We use \emph{syntactic validation} broadly to mean validation enforced at the boundary against a specified representation or verification rule. Signature validation, certificate validation, and authentication-token validation are syntactic in this sense even when they use cryptographic mechanisms. They establish assertions such as whether an artifact verifies under an accepted key, whether a certificate satisfies configured issuer and lifetime rules, or whether a token is well formed, properly signed, and consistent with expected fields. They do not by themselves establish that the verified principal is the principal the receiving domain intends to trust, that the artifact is safe to execute, that the requested operation is within the intended authority, or that the assertion will remain valid when later used by the receiving domain.

\begin{definition}[Trust Boundary Semantic Gap]
\label{def:tbsg}
A TBSG exists at $TB_{i,j}$ for an artifact $A$ when $A$ crosses the boundary, passes the syntactic validation applied by the receiving domain, and the assertions established by that pass do not entail the security requirements that $D_j$ places on $A$:
\[
  \Phi(A) \not\models R_j(A).
\]
\end{definition}

A TBSG is not a failure of syntactic validation. The gap lies between what syntactic validation establishes, $\Phi(A)$, and what the receiving domain needs to process artifact $A$ safely, $R_j(A)$. We use four assumptions to scope TBSG analysis. First, the syntactic specification enforced at the boundary is correctly implemented; otherwise, the case is an implementation flaw. Second, the specification reflects what the boundary is designed to enforce; if an intended syntactic requirement is missing from it, the case is a specification error. Third, TBSG analysis applies only when syntactic validation exists and the artifact passes it. Fourth, TBSG is evaluated per artifact at each trust boundary, because different artifacts may carry different assertions and require different security requirements. A TBSG is evaluated per boundary, but when $D_j$ processes $A$ despite $\Phi(A) \not\models R_j(A)$, artifacts that $D_j$ later emits may carry the unverified assertion into subsequent boundaries, where validation succeeds because $D_j$ is an accepted source. These definitions provide the basis for TBSAM in Section~\ref{sec:tbsam}: Definition~\ref{def:trust_boundary} guides trust boundary identification, and the comparison between $\Phi(A)$ and $R_j(A)$ guides TBSG identification and dimension analysis. 


\subsection{Four Dimensions of Semantic Misalignment}
\label{subsec:MDTBSG}
The incidents analyzed in Section~\ref{subsec:incident_analysis} show that TBSG does not appear in a single fixed form. We organize the recurring forms of semantic misalignment into four dimensions, related to what the receiving domain still lacks after syntactic validation succeeds: \textit{Identity} (\textit{who} is behind the artifact), \textit{Spatial}  (\textit{where} the artifact reaches), \textit{Temporal} (\textit{when} an assertion remains valid), and \textit{Interpretation} (\textit{what} behavior the artifact causes).

Each dimension follows the definition explained in Section~\ref{subsec:tbs_definition}: syntactic validation establishes $\Phi(A)$, but $\Phi(A) \not\models R_j^{d}(A)$ for the corresponding class. The dimensions differ only in what $\Phi(A)$ establishes and what the
receiving domain still needs, so we describe each dimension using these two parts.

\subsubsection{Identity Dimension}
\label{sec:dim-identity}
$\Phi(A)$ establishes the structural validity of a credential, token, certificate, or signing key. $R_j^{\mathrm{Id}}(A)$ requires assurance that the actor behind the artifact is the principal the receiving domain intends to trust. A structurally valid credential may be presented by an actor who should not be trusted as the corresponding principal, or an actor may accumulate trust over time and later use that position for malicious purposes. Confused-deputy scenarios in which a privileged component acts on behalf of an unverified requester fall partly in this dimension and partly in the Spatial dimension, depending on whether the missing assurance concerns the actor or the reach of the action~\cite{hardy1988confused}.

\subsubsection{Spatial Dimension}
\label{sec:dim-spatial}
$\Phi(A)$ confirms the well-formedness of a destination identifier, such as a URL, domain name, IP address, path, or policy field. $R_j^{\mathrm{Sp}}(A)$ requires that the artifact reaches only the intended destination and that its effects remain within the expected scope. A well-formed destination identifier may resolve to an internal service the receiving domain did not intend to expose, or a valid path may traverse beyond the directory boundary the receiver assumes.

\subsubsection{Temporal Dimension}
\label{sec:dim-temporal}
$\Phi(A)$ establishes the validity of an assertion at validation time. $R_j^{\mathrm{Te}}(A)$ requires that the same assertion remains valid at use time. Syntactic validation observes a point-in-time state and does not capture later changes in system state, principal status, authorization context, or artifact freshness. This dimension subsumes time-of-check-to-time-of-use (TOCTOU) conditions, classically studied as file-system races~\cite{bishop1996race}, and extends the same pattern to any assertion whose validity changes between validation and use. Remote-attestation results show the same structure when a measured state changes between attestation and use~\cite{nunes2021toctou}.

\subsubsection{Interpretation Dimension}
\label{sec:dim-interpretation}

$\Phi(A)$ establishes that an artifact conforms to an expected schema, grammar, or syntactic structure. $R_j^{\mathrm{In}}(A)$ requires that the artifact produces only the behavior the receiving domain intends. A syntactically valid artifact may trigger behavior that differs from the meaning the receiver assumes, as when input intended as data is processed as executable code or a command. This dimension subsumes injection-style conditions and the parser differentials studied in language-theoretic security
(LangSec)~\cite{sassaman2013security}.

\subsubsection{MDTBSG as a Four-Dimensional Analysis Model}
\label{sec:MDTBSG_model}

MDTBSG is a four-dimensional analysis model of semantic
misalignment at trust boundaries. It asks a different question from attack-oriented classifications. STRIDE enumerates classes of adversarial action against a system element, such as spoofing, tampering, repudiation, information disclosure, denial of service, and elevation of privilege. MDTBSG instead classifies, at a single trust boundary, what the receiving domain still lacks after syntactic validation succeeds: the unestablished security property, not the adversarial behavior.

This distinction is not about which classification is more
correct. It is about the property each classification captures. Spoofing, for example, names an adversarial action. The Identity dimension names the boundary-level condition that can let the action succeed: a structurally valid credential that does not establish the actor behind it. Because the two classifications describe different aspects of the same boundary crossing, they do not align one-to-one. A single STRIDE threat may involve gaps in multiple MDTBSG dimensions, and a single MDTBSG dimension may appear through different STRIDE categories.

TBSAM uses the four dimensions to ask, for each artifact and each trust boundary, which security property remains unestablished after syntactic validation succeeds. The same dimensions also provide shared labels for tracing how a gap at one boundary appears again at a subsequent boundary.

Existing weakness vocabularies such as CWE capture TBSG
only partially. A single trust-boundary semantic gap may span several CWE entries, and one CWE entry may appear across different MDTBSG dimensions. This many-to-many relation exists because CWE classifies the software weakness, whereas MDTBSG classifies the security property that the receiving domain leaves unestablished at the boundary.

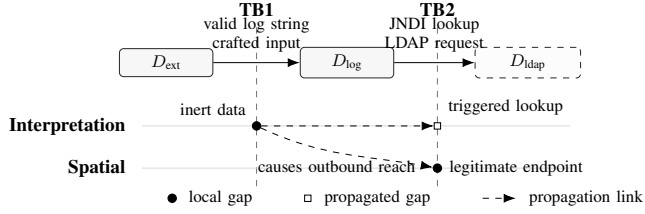
\begin{figure}[t]
\centering
\resizebox{\columnwidth}{!}{%
\begin{tikzpicture}[
x=1cm, y=1cm,
font=\footnotesize,
domainbox/.style={
draw=black!85,
rounded corners=2pt,
minimum width=1.55cm,
minimum height=0.45cm,
align=center,
fill=black!3
},
attacker/.style={
draw=black!85,
rounded corners=2pt,
minimum width=1.75cm,
minimum height=0.45cm,
align=center,
dashed,
fill=black!1
},
boundary/.style={
dashed,
line width=0.4pt,
black!55
},
flow/.style={
-{Latex[length=2mm]},
line width=0.45pt
},
prop/.style={
-{Latex[length=2mm]},
dashed,
line width=0.45pt
},
pOne/.style={
circle,
draw=black,
fill=black,
inner sep=1.4pt
},
pThree/.style={
rectangle,
draw=black,
fill=white,
inner sep=1.5pt
}
]

\node[domainbox] (ext) at (0,2.3) {$D_{\text{ext}}$};
\node[domainbox] (log) at (3.0,2.3) {$D_{\text{log}}$};
\node[attacker] (ldap) at (6.0,2.3) {$D_{\text{ldap}}$};

\draw[boundary] (1.5,0.25) -- (1.5,2.95);
\draw[boundary] (4.5,0.25) -- (4.5,2.95);

\node[font=\small\bfseries] at (1.5,3.18) {TB1};
\node[font=\small\bfseries] at (4.5,3.18) {TB2};

\draw[flow] (ext) -- node[above=1.5pt, font=\footnotesize, align=center]
{valid log string\\crafted input} (log);

\draw[flow] (log) -- node[above=1.5pt, font=\footnotesize, align=center]
{JNDI lookup\\LDAP request} (ldap);

\draw[black!15] (-0.4,1.25) -- (6.7,1.25);
\draw[black!15] (-0.4,0.55) -- (6.7,0.55);

\node[anchor=east, font=\small\bfseries] at (-0.55,1.25) {Interpretation};
\node[anchor=east, font=\small\bfseries] at (-0.55,0.55) {Spatial};

\node[pOne, label={[font=\footnotesize]above left:{inert data}}] (i1) at (1.5,1.25) {};
\node[pThree, label={[font=\footnotesize]above right:{triggered lookup}}] (i2) at (4.5,1.25) {};

\node[pOne, label={[font=\footnotesize]right:{legitimate endpoint}}] (s1) at (4.5,0.55) {};

\draw[prop] (i1) -- (i2);
\draw[prop] (i1) to[out=-25,in=170] node[below, pos=0.45, font=\footnotesize] {causes outbound reach} (s1);

\node[pOne] at (0.1,0.05) {};
\node[anchor=west, font=\footnotesize] at (0.25,0.05) {local gap};

\node[pThree] at (2.35,0.05) {};
\node[anchor=west, font=\footnotesize] at (2.5,0.05) {propagated gap};

\draw[prop] (5.25,0.05) -- (5.8,0.05);
\node[anchor=west, font=\footnotesize] at (5.9,0.05) {propagation link};

\end{tikzpicture}%
}
\caption{Log4Shell boundary-level propagation. A syntactically valid log string first creates an Interpretation gap at TB1 because validation does not establish that logged input remains inert data. The resulting JNDI lookup then creates a Spatial gap at TB2 because a well-formed destination does not establish that the endpoint is legitimate.}
\label{fig:log4shell_propagation}
\end{figure}

Log4Shell~\cite{nvd_cve202144228} illustrates how one incident can span more
than one dimension and boundary. Fig.~\ref{fig:log4shell_propagation} shows how Interpretation-class gap at the first boundary leads to a Spatial-class gap at the subsequent boundary. An attacker supplies a string, such as a crafted HTTP header, that crosses from an external client ($D_i$) into a logging component ($D_j$). The logger parses the string as a valid log message, so $\Phi(A)$ establishes structural conformance. The receiving domain, however, requires logged input to remain inert data, an Interpretation-class property in $R_j^{\mathrm{In}}(A)$. This property does not follow from $\Phi(A)$; hence
$\Phi(A) \not\models R_j^{\mathrm{In}}(A)$, and the TBSG appears in the Interpretation dimension.

Because this gap remains open, the embedded JNDI lookup causes $D_j$ to emit an outbound request. That request contains a well-formed destination identifier, but the identifier resolves to attacker-controlled infrastructure rather than an intended endpoint. The subsequent boundary therefore exposes a Spatial-class gap. A single incident thus requires per-artifact, per-boundary analysis and may involve different MDTBSG dimensions at different points in the exploit path.

\section{Trust Boundary Semantic Analysis and Mitigation}
\label{sec:tbsam}

Section~\ref{sec:tbsg} defines TBSG and organizes semantic gaps into the four MDTBSG dimensions. This section presents Trust Boundary Semantic Analysis and Mitigation (TBSAM), a design-time framework composed of a four-stage procedure. TBSAM applies those dimensions to system design artifacts, identifies semantic gaps at trust boundaries, and maps them to architectural controls. Between prioritization and mitigation, propagation analysis (Section~\ref{subsec:propagation}) traces how an unresolved gap at one boundary can appear again when a later receiving domain trusts an assertion that earlier validation did not establish.

\subsection{Overview of TBSAM}
\label{subsec:overview_TBSAM}
Syntactic validation is necessary as the first line of defense, but it does not by itself guarantee the receiving domain's security properties. TBSAM makes this mismatch explicit before implementation. It helps security architects compare what syntactic validation establishes at a trust boundary with what the receiving domain requires for safe processing.

TBSAM takes system design artifacts, such as data flow diagrams, architecture documents, and API specifications, as input. It produces a prioritized list of gaps mapped to architectural controls. As a design-time procedure, TBSAM runs in parallel with the threat modeling that designers already perform. Threat modeling with STRIDE proceeds in three steps: draw the data-flow diagram, enumerate threats at each element or boundary, and map each threat to a control. TBSAM follows a similar path on the same design artifacts: identify the trust boundary, record the scope of syntactic validation, derive the unestablished properties as an MDTBSG dimension, and map it to an architectural control. It begins after a threat-modeling control has been selected. 
 
TBSAM is intended as a design-review procedure rather than a fully automated analysis or runtime detector. TBSAM can be applied after an architectural control has been selected during STRIDE-based threat modeling. For example, code signing may be selected as a mitigation for tampering. From TBSAM's point of view, that control becomes part of the boundary's syntactic validation $\Phi(A)$. TBSAM then asks a residual question: \textit{after this control succeeds, does the receiving domain still rely on a property the control did not establish?} A valid signature establishes artifact integrity and signer identity, but not that the build process behind the signature was uncompromised. Used this way, TBSAM does not duplicate threat enumeration. It takes the selected architectural control as its starting point and examines what remains unestablished at the boundary.
 
\subsection{TBSAM Procedure}
\label{subsec:tbsam_procedure}
TBSAM uses a four-stage procedure. Stage~1 constructs a boundary record for each confirmed trust boundary. Stage~2 uses the record to select candidate MDTBSG dimensions. Stage~3 assigns priorities to the candidate gaps. Propagation analysis then traces each P3 gap to the preceding boundary where it originates (Section~\ref{subsec:propagation}). Stage~4 maps the prioritized gaps to architectural controls. 

\subsubsection{Stage 1: Boundary Record Construction}
\label{subsubsec:stage1_boundary_record}

Design documents rarely mark trust boundaries explicitly. Stage~1 identifies trust boundaries from design artifacts and records the syntactic validation performed at each boundary. The security architect first looks for points where the security policy changes. Such changes appear when the authentication or authorization subject changes, the privilege model changes, or trust assumptions differ between two domains. Points without these signals are excluded. For each remaining point, the security architect checks whether an artifact crosses domains and whether the receiving domain performs syntactic validation on that artifact. Points without artifact movement are excluded. Points with artifact movement but without syntactic validation are recorded separately as missing syntactic validation, because TBSG assumes that syntactic validation exists and is passed. For each confirmed trust boundary, the boundary record stores three items: the crossing artifact, the syntactic validation performed by the receiving domain and the properties it covers, and whether the current design treats the artifact as passing syntactic validation. This record becomes the input to Stage~2.

\subsubsection{Stage 2: Candidate Dimension Selection from Syntactic Scope}
\label{subsubsec:stage2_candidate_dimension}

Stage~2 compares the assertions established by syntactic validation, $\Phi(A)$, with the security properties that the receiving domain
places on the artifact, $R_j(A)$. TBSAM records dimensions in which a gap may exist as candidate dimensions. The comparison uses the scope of syntactic validation from the boundary record and the security property required by the receiving domain. Because the design document often does not state this property directly, the security architect identifies it from the receiving domain's security policy, subsequent security-sensitive action and the assumptions made by syntactic validation. The security architect compares the reconstructed properties with the scope of syntactic validation across the four MDTBSG dimensions: Identity, Spatial, Temporal, and Interpretation. If a possible gap is found, the corresponding dimension is added to the candidate dimension set. This set records possible gaps only; Stage~3 assesses exploitability and threat relevance.

\subsubsection{Stage 3: Dimension Refinement via Priority Filtering}
\label{subsubsec:stage3_priority_filtering}

Stage~3 refines the candidate dimension set and assigns each selected gap to P1, P2, or P3. The goal is to decide which gaps should be mitigated first. The security architect first identifies the primary security property of the boundary. This is the property that must hold before the receiving domain can safely perform its first security-sensitive action on the accepted artifact. The property is inferred from the boundary record, the scope of syntactic validation, and the downstream behavior of the receiving domain. The three priority levels are defined as follows. P1 marks a gap that originates at the current boundary and directly violates the receiving domain's primary security property; mitigating it is necessary to interrupt the attacker's objective at that boundary. P2 marks a gap that is not the primary gap to be mitigated first, but enables, amplifies, stabilizes, or extends the effect of a P1 gap. P3 marks a gap that is visible at the current boundary but originates from an unresolved gap at a preceding boundary; propagation analysis (Section~4.2.4) identifies the preceding P1 source. The output of Stage 3 is a prioritized list of dimension gaps.

\subsubsection{Propagation Analysis}
\label{subsec:propagation}

A P3 gap marks a propagated condition. It appears at the current boundary because a preceding boundary $TB_{i,j}$ left a semantic gap unresolved. Propagation analysis makes this chain explicit. Its goal is to place the control at the boundary where the gap starts, not only at the later boundary where the gap becomes visible. TBSAM traces this chain across consecutive trust boundaries using Definitions~\ref{def:trust_boundary} and~\ref{def:tbsg}. The four dimensions provide the common coordinate for this step. For each P3 gap, the security architect identifies the preceding boundary where the same semantic misalignment first appears as P1. The architect then checks whether the P1 control at that boundary removes the unverified assertion before the artifact reaches the subsequent boundary. If it does not, the current boundary treats the gap as P1. This step prevents a local-only view of trust boundaries. A subsequent boundary may accept a syntactically valid artifact while still relying on an unverified assertion created at a preceding boundary.

\subsubsection{Stage 4: Mitigation Mapping}
\label{subsubsec:stage4}

Stage~4 maps each prioritized gap to architectural controls. The security architect handles P1 gaps first, because they mark the boundary where the semantic gap starts. P2 gaps receive additional controls when they remain relevant to the receiving domain's security property. P3 markers are not handled as independent local gaps; TBSAM traces them to the preceding boundary where the same gap first appears as P1. The output of this stage is a set of architectural controls for each trust boundary. The controls should make the assertion established by syntactic validation closer to the security property required by the receiving domain.

\begin{table*}[t]
\centering
\caption{TBSAM Dimension–Candidate Mitigation Mapping}
\label{tab:mitigation_detailed}
\footnotesize
\renewcommand{\arraystretch}{1.12}
\begin{tabularx}{\textwidth}{p{1.8cm}p{3.0cm}p{4.8cm}X}
\toprule
\textbf{Dimension} &
\textbf{Gap Target} &
\textbf{Mitigation Strategy} &
\textbf{Example Technique} \\
\midrule

Identity
  & Credential, token, service identity, long-lived trust
  & Bind identity assertions to the environment, session, workload, or current trust state.
  & TPM-backed attestation, token binding, SPIFFE/SPIRE workload identity, continuous access evaluation. \\

Spatial
  & URL, file path, network destination, cloud resource scope
  & Restrict where an artifact can reach and where its effects can occur.
  & SSRF allowlists, resolved-IP checks, path canonicalization, eBPF-based egress policy, per-request scope checks. \\

Temporal
  & Session token, file-system state, authorization context, artifact freshness
  & Reduce or bind the time window between check and use.
  & Short-lived tokens, atomic check-use binding, behavioral heartbeat, signed timestamps, provenance freshness checks. \\

Interpretation
  & Build artifact, input data, software component, dynamic loading
  & Ensure that a syntactically valid artifact preserves its intended security meaning when processed.
  & Source-to-binary attestation, reproducible builds, runtime behavior policy enforcement, SBOM or in-toto checks, resolution rule hardening. \\

\bottomrule
\end{tabularx}
\end{table*}

\section{Case Study: SolarWinds/SUNBURST}
\label{sec:casestudy}

This section applies TBSAM to the SolarWinds/SUNBURST supply-chain attack.
The case study has two goals. First, Section~\ref{subsec:applying_tbsam} shows that the four-stage procedure in Section~\ref{sec:tbsam} applies to a real system attack.
Second, Section~\ref{subsec:crossboundary} shows why cross-boundary
propagation matters. The SolarWinds exploit path was not a single isolated
failure but a chain of TBSGs across multiple trust boundaries. TBSAM records
this chain through P3 markers.

\subsection{Analysis Target and Input}
\label{subsec:target_input}

SolarWinds Orion is a network monitoring platform used in enterprise and
government environments. Orion updates distributed between March and June
2020 contained the SUNBURST backdoor. The threat actor inserted the backdoor
after compromising the SolarWinds build environment and changing the build
process. About 18,000 organizations installed the affected updates. This case is a useful stress test for design-level analysis. At each boundary
crossed by the attack, the deployed syntactic validation mechanisms worked as
designed. The actor authenticated with valid credentials. The modified DLL
compiled without error and carried a valid SolarWinds Authenticode signature.
Customers downloaded the package from the official channel and checked its
signature and hash. The backdoor's outbound traffic also conformed to DNS,
TLS, and HTTP protocol checks. The attack therefore did not proceed by
bypassing syntactic validation. It proceeded by satisfying it.

This is the condition captured by TBSG. At each boundary,
syntactic validation
established $\Phi(A)$, but the security properties required by the receiving
domain remained unestablished:
\[
    \Phi(A) \not\models R_j(A).
\]

SolarWinds has not published the internal design of its build and distribution
pipeline. We therefore reconstruct the trust domains, trust boundaries, and
syntactic validation mechanisms from public post-incident reports:
CrowdStrike's SUNSPOT analysis~\cite{crowdstrike_sunspot}, Mandiant's
SUNBURST analysis~\cite{mandiant_sunburst}, ReversingLabs' build-process
analysis~\cite{reversinglabs_sunburst}, and CISA Alert
AA20-352A~\cite{cisa_aa20352a}. The reconstruction uses only design-level
facts that these reports establish independently of the attack outcome:
which domains existed, which boundaries connected them, and which syntactic
validation mechanisms the receiving domains used.

We use the same reconstructed model as input for both analyses in this section: TBSAM in Section~\ref{subsec:applying_tbsam} and STRIDE in
Section~\ref{subsubsec:stride_comparison}. 

\subsection{Applying TBSAM}
\label{subsec:applying_tbsam}

\subsubsection{Stage 1: Boundary Record Construction}
\label{subsubsec:sw_stage1}

Following the system structure described in the public reports, Stage~1 identifies five trust domains: the external environment ($D_{\text{ext}}$), the build server ($D_{\text{build}}$), the distribution server ($D_{\text{dist}}$), the customer system ($D_{\text{cust}}$), and the attacker infrastructure ($D_{\text{c2}}$). These domains operate under different security assumptions and form four trust boundaries (Fig.~\ref{fig:propagation}). Table~\ref{tab:sw_boundary_record} summarizes the boundary record, and Appendix~\ref{app:solarwinds_boundary_record} gives the full record.

The attack path begins at \textbf{TB1} ($D_{\text{ext}} \to D_{\text{build}}$) where the external
environment reaches the build server through remote-access and credential-based entry points. It then crosses TB2 ($D_{\text{build}} \to D_{\text{dist}}$), where build output
moves from the compilation environment into the distribution path.
At this boundary, the modified DLL was accepted as a valid SolarWinds
artifact because it compiled successfully and carried a valid
Authenticode signature~\cite{reversinglabs_sunburst}.

The signed package next crosses TB3 ($D_{\text{dist}} \to D_{\text{cust}}$), where customer systems retrieve the update through the official channel and verify its signature and package integrity. The final boundary, TB4 ($D_{\text{cust}} \to D_{\text{c2}}$), covers outbound communication from the infected Orion instance to attacker infrastructure. At this boundary, SUNBURST's DNS and HTTPS traffic conformed to expected protocol checks and was disguised as normal Orion communication using the Orion Improvement Program protocol~\cite{mandiant_sunburst}.

\subsubsection{Stage 2: Candidate Dimension Selection}
\label{subsubsec:sw_stage2}

For each boundary, Stage~2 compares the assertion established by syntactic
validation, $\Phi(A)$, with the security property required by the receiving
domain, written as $R_j(A)$ in the formal definition. Stage~2 selects a
candidate dimension when the established assertion does not satisfy that
property, i.e., when $\Phi(A) \not\models R_j(A)$.

\textbf{TB1} selects \emph{Identity}. Credential and session checks establish
that the submitted credential and session are valid. They do not establish
that the actor using the credential is an authorized SolarWinds developer.
The threat actor crossed TB1 with valid credentials.

\textbf{TB2} selects \emph{Interpretation} and \emph{Temporal}. Code signing
and successful compilation establish that the binary has the expected
representation, not that the binary performs only the behavior SolarWinds
intends (Interpretation). CrowdStrike reports that SUNSPOT monitored the
MSBuild process, replaced source files immediately before compilation, and
performed consistency checks to avoid build errors~\cite{crowdstrike_sunspot}.
As a result, the assertion ``compiled successfully and signed correctly''
held for a modified artifact. TB2 also carries a Temporal candidate because
the build process checks the artifact at one point but uses it later as
trusted build output.

\textbf{TB3} selects \emph{Interpretation}, \emph{Spatial}, and
\emph{Temporal}. Customer-side Authenticode verification establishes the same
assertion that was already insufficient at TB2: the package is correctly
signed, but its intended behavior remains unestablished (Interpretation). The
official update-channel check shows that the package came from the expected
distribution location, not that the full supply-chain path was uncompromised
(Spatial). Hash comparison confirms integrity at download time, not across
the interval from build to customer download (Temporal).

\textbf{TB4} selects \emph{Spatial} and \emph{Temporal}. DNS responses and
TLS certificates establish protocol conformance of the outbound channel, not
that the destination is a legitimate endpoint for Orion communication
(Spatial). SUNBURST's 12--14 day dormancy after first execution also moved
its malicious behavior outside the time window covered by earlier checks
(Temporal).

\subsubsection{Stage 3: Priority Assignment}
\label{subsubsec:sw_stage3}

Stage~3 applies the priority rules from
Section~\ref{subsubsec:stage3_priority_filtering} to the Stage~2 candidates.
Each assignment below includes a short rationale based on those rules. This
makes the assignment checkable from the boundary record, rather than dependent
on a free-form reading of the incident.

\textbf{TB1 / Identity $\to$ P1.} The Identity gap starts at TB1 and has no
preceding source. If the entry boundary accepts the actor as a legitimate
build-environment user, subsequent boundaries inherit that trust decision.
This makes TB1's Identity gap a root condition for the exploit path.

\textbf{TB2 / Interpretation $\to$ P1.} The Interpretation gap starts at TB2.
No preceding boundary can establish the semantic behavior of the binary
because the binary does not exist before the build process produces it. TB2
is therefore the first boundary that can check whether the generated artifact
matches the intended source-level behavior.

\textbf{TB2 / Temporal $\to$ P2.} The Temporal gap is relevant, but
mitigating TB2's P1 gap reduces its effect. Source-to-binary attestation ties
the artifact's later use back to the build steps that produced it, narrowing
the check-time/use-time window available to the attacker.

\textbf{TB3 / Interpretation $\to$ P3.} The Interpretation gap at TB3 is
inherited from TB2. Customer-side signature verification re-establishes the
same assertion over the same signed artifact. The propagation rule marks this
gap P3 with a pointer to TB2's P1 gap. Closing TB2's P1 gap also closes this
propagated gap.

\textbf{TB3 / Spatial $\to$ P2, TB3 / Temporal $\to$ P2.} Both gaps are
secondary. Their impact depends on the fact that TB3 has already accepted the
malicious artifact as a legitimate SolarWinds update. Applying TB2's P1
control across the supply chain reduces both gaps at TB3.

\textbf{TB4 / Spatial $\to$ P1.} The Spatial gap starts locally at TB4 and
has no preceding source. Even if TB1--TB3 were fully mitigated, preceding
boundaries still would not establish whether an outbound destination is
legitimate for Orion communication. The outbound communication boundary must
check this property itself.

\textbf{TB4 / Temporal $\to$ P2.} Dormancy widens the time gap between initial
checks and malicious behavior, but its effect depends on the Spatial gap at
TB4. If TB4 checks destination legitimacy, the dormancy window alone does not
give the attacker a useful outbound path.

\subsubsection{Stage 4: Mitigation Mapping}
\label{subsubsec:sw_stage4}

Stage~4 maps each prioritized gap to architectural controls using
Table~\ref{tab:mitigation_detailed}. For \textbf{TB1's P1 Identity} gap,
TBSAM maps attestation-based identity and continuous re-authentication, so
that the build server reasons about the accessing subject rather than only
the well-formedness of a credential. For \textbf{TB3's propagated
Interpretation} gap (P3), the customer side needs no independent control of
the same strength: applying TB2's control across the supply chain closes the
gap, with SBOM-based component checking as defense in depth. For
\textbf{TB4's P1 Spatial} gap, TBSAM maps semantic micro-segmentation and
destination checks: the control restricts outbound communication from Orion
instances to the legitimate SolarWinds infrastructure set, which blocks
SUNBURST's DGA-generated DNS queries.

The \textbf{TB2 P1 Interpretation} mapping captures the central point of the
paper. Stronger signature verification at TB2 cannot close this gap, because
the signing step lies inside the compromised generation process; re-checking
the signature only confirms the representation the attacker preserved. The
control must instead check the generation context---whether each step from
source commit to signed binary produced output within the expected
range---which is what source-to-binary attestation and reproducible builds
provide. Under reproducible builds, SUNSPOT's pre-compilation source
substitution would yield a binary that an independent rebuild does not
reproduce, exposing the modification at the boundary where the gap originates. Under this reconstruction, the boundary record alone — without SUNBURST-specific indicators — identifies TB2 as the origin of the Interpretation gap; a system with the same boundary structure would therefore receive the same Stage~4 control mapping.

\subsection{Cross-Boundary Analysis}
\label{subsec:crossboundary}

\begin{figure*}[!t]
\centering
\begin{tikzpicture}[
  x=1cm, y=1cm,
  font=\small,
  domnode/.style={
    draw=black!80, line width=0.55pt, rounded corners=2pt,
    minimum width=1.65cm, minimum height=0.60cm,
    align=center, fill=black!3, font=\footnotesize\bfseries
  },
  atkdom/.style={
    domnode, dashed, fill=white
  },
  patharrow/.style={
    -{Stealth[length=2.0mm,width=1.6mm]},
    line width=0.65pt, draw=black!85
  },
  proparrow/.style={
    -{Stealth[length=1.8mm,width=1.4mm]},
    dashed, line width=0.55pt, draw=black!75
  },
  carryarrow/.style={
    -{Stealth[length=1.7mm,width=1.3mm]},
    dotted, line width=0.65pt, draw=black!65
  },
  condarrow/.style={
    -{Stealth[length=1.7mm,width=1.3mm]},
    densely dashed, line width=0.55pt, draw=black!65
  },
  tbline/.style={
    dashed, line width=0.5pt, draw=black!30
  },
  guideln/.style={
    line width=0.45pt, draw=black!10
  },
  lanelabel/.style={
    font=\footnotesize\bfseries, anchor=east, text=black!85
  },
  note/.style={
    font=\tiny, text=black!85, align=center
  },
  tag/.style={
    font=\scriptsize\bfseries, text=black!85,
    fill=white, inner sep=1pt
  },
  edgelabel/.style={
    font=\tiny\itshape, text=black!75,
    fill=white, inner sep=1pt
  },
  p1/.style={
    circle, draw=black!80, fill=black!80,
    minimum size=5.0pt, inner sep=0pt
  },
  rootring/.style={
    circle, draw=black!80, line width=0.65pt,
    minimum size=9.2pt, inner sep=0pt
  },
  p2/.style={
    diamond, draw=black!80, fill=white, line width=0.65pt,
    minimum size=6.4pt, inner sep=0pt
  },
  p3/.style={
    rectangle, draw=black!80, fill=black!10, line width=0.65pt,
    minimum size=5.6pt, inner sep=0pt
  }
]

\def\xa{0.7}
\def\xb{3.4}
\def\xc{6.1}
\def\xd{8.8}
\def\xe{11.5}

\def\tbA{2.05}
\def\tbB{4.75}
\def\tbC{7.45}
\def\tbD{10.15}

\def\yId{-1.55}
\def\yIn{-2.70}
\def\yTe{-4.05}
\def\ySp{-5.25}

\node[domnode] (ext)   at (\xa,0) {$D_{\mathrm{ext}}$};
\node[domnode] (build) at (\xb,0) {$D_{\mathrm{build}}$};
\node[domnode] (dist)  at (\xc,0) {$D_{\mathrm{dist}}$};
\node[domnode] (cust)  at (\xd,0) {$D_{\mathrm{cust}}$};
\node[atkdom]  (cc)    at (\xe,0) {$D_{\mathrm{c2}}$};

\draw[patharrow] (ext) -- (build);
\draw[patharrow] (build) -- (dist);
\draw[patharrow] (dist) -- (cust);
\draw[patharrow] (cust) -- (cc);

\node[font=\scriptsize\itshape\bfseries, anchor=east, text=black!55]
  at (-0.15,0) {Exploit path};

\foreach \x/\n in {\tbA/TB1,\tbB/TB2,\tbC/TB3,\tbD/TB4}{
  \draw[tbline] (\x,0.48) -- (\x,-5.88);
  \node[tag, text=black!55] at (\x,0.70) {\n};
}

\node[lanelabel] at (-0.15,\yId) {Identity};
\node[lanelabel] at (-0.15,\yIn) {Interpretation};
\node[lanelabel] at (-0.15,\yTe) {Temporal};
\node[lanelabel] at (-0.15,\ySp) {Spatial};

\foreach \y in {-0.95,-2.10,-3.35,-4.65,-5.88}{
  \draw[guideln] (0.05,\y) -- (12.15,\y);
}

\node[note, fill=white, inner sep=1.5pt] at (\tbA,-0.55)
  {remote-access\\session / credential};

\node[note, fill=white, inner sep=1.5pt] at (\tbB,-0.55)
  {signed build\\artifact (.dll)};

\node[note, fill=white, inner sep=1.5pt] at (\tbC,-0.55)
  {signed update\\package (.msp)};

\node[note, fill=white, inner sep=1.5pt] at (\tbD,-0.55)
  {DNS query +\\HTTPS beacon};

\node[rootring] at (\tbA,\yId) {};
\node[p1] (idroot) at (\tbA,\yId) {};
\node[tag, anchor=east] at (\tbA-0.22,\yId)
  {P1 root};

\draw[carryarrow] (\tbA+0.16,\yId) -- (\xe-0.25,\yId)
  node[midway, above, edgelabel]
  {carried into subsequent artifacts};

\node[p1] (inp1) at (\tbB,\yIn) {};
\node[tag, anchor=south] at (\tbB,\yIn+0.12)
  {P1};

\node[p3] (inp3) at (\tbC,\yIn) {};
\node[tag, anchor=west] at (\tbC+0.16,\yIn)
  {P3};

\node[note, anchor=west, text=black!70] at (\tbC+0.55,\yIn-0.22)
  {closed by\\TB2 control};

\draw[proparrow] (inp1) -- (inp3)
  node[midway, above, edgelabel]
  {propagation};

\draw[condarrow] (idroot.south east) to[out=-35, in=145]
  node[pos=0.55, above, sloped, edgelabel]
  {conditions}
  ($(inp1)+(-0.13,0.13)$);

\fill[black!5]
  (\tbB,\yTe) -- (\tbD+0.35,\yTe+0.24) --
  (\tbD+0.35,\yTe-0.24) -- cycle;

\draw[black!55, line width=0.5pt]
  (\tbB,\yTe) -- (\tbD+0.35,\yTe+0.24) --
  (\tbD+0.35,\yTe-0.24) -- cycle;

\node[p2] at (\tbB,\yTe) {};
\node[p2] at (\tbC,\yTe) {};
\node[p2] at (\tbD,\yTe) {};

\foreach \x in {\tbB,\tbC,\tbD}{
  \node[tag, anchor=north] at (\x,\yTe-0.22) {P2};
}

\node[edgelabel, anchor=south] at (\tbC+1.10,\yTe+0.32)
  {TOCTOU window widens};

\node[p2] at (\tbC,\ySp) {};
\node[tag, anchor=north] at (\tbC,\ySp-0.18)
  {P2};

\node[p1] at (\tbD,\ySp) {};
\node[tag, anchor=north, align=center] at (\tbD,\ySp-0.18)
  {P1\\local};

\begin{scope}[shift={(0.05,-6.70)}]
  \draw[rounded corners=2pt, draw=black!20, fill=black!1]
    (-0.25,0.36) rectangle (12.25,-0.36);

  \node[rootring] at (0,0) {};
  \node[p1] at (0,0) {};
  \node[note, anchor=west] at (0.22,0)
    {root gap (P1)};

  \node[p1] at (2.35,0) {};
  \node[note, anchor=west] at (2.55,0)
    {local gap (P1)};

  \node[p2] at (4.75,0) {};
  \node[note, anchor=west] at (4.98,0)
    {secondary gap (P2)};

  \node[p3] at (7.25,0) {};
  \node[note, anchor=west] at (7.48,0)
    {propagated gap (P3)};

  \draw[proparrow] (9.80,0) -- (10.40,0);
  \node[note, anchor=west] at (10.55,0)
    {propagation link};
\end{scope}

\end{tikzpicture}

\caption{SolarWinds/SUNBURST cross-boundary propagation graph. The top lane
shows the exploit path across five trust domains and four trust boundaries.
The lower lanes show where Identity, Interpretation, Temporal, and Spatial
gaps originate, propagate, or amplify. Marker shape and line style encode the
gap role, making the graph safe for grayscale printing.}
\label{fig:propagation}
\end{figure*}
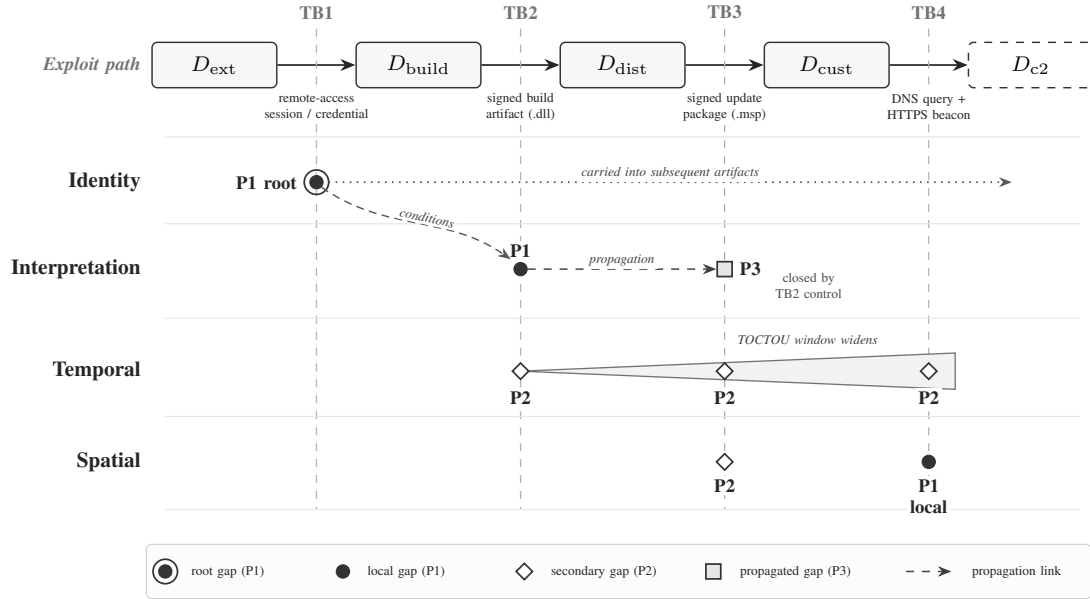

\begin{table*}[t]
\caption{Positioning of TBSAM relative to existing security analysis
approaches. TBSAM starts from a successful syntactic validation pass and asks
which receiving-domain security requirements remain unestablished across trust
boundaries.}
\label{tab:positioning}
\centering
\footnotesize
\setlength{\tabcolsep}{4pt}
\renewcommand{\arraystretch}{1.18}

\begin{tabularx}{\textwidth}{@{}
>{\raggedright\arraybackslash}p{0.18\textwidth}
>{\raggedright\arraybackslash}X
>{\raggedright\arraybackslash}p{0.21\textwidth}
>{\raggedright\arraybackslash}X
@{}}
\toprule
\textbf{Approach} &
\textbf{Starting question} &
\textbf{Finding} &
\textbf{Propagation view} \\
\midrule

STRIDE-based threat modeling~\cite{kohnfelder1999threats,shostack2014threat} &
Which threat categories apply to each DFD element or flow? &
Threat category mapped to a control. &
Findings are organized per DFD element, flow, or boundary. Each finding is
handled where it appears. \\

\addlinespace[2pt]

SLSA~\cite{slsa} &
Which supply-chain integrity level does the build and release pipeline meet? &
Controls required for a target integrity level. &
Findings follow a per-level checklist. Boundary structure is implicit in the
pipeline stages. \\

\addlinespace[2pt]

Semantic-gap analyses~\cite{garfinkel2003vmi,fu2012space,jain2014sok,
su2006essence} &
Does meaning on one side match the meaning required on the other side in the
studied setting? &
Setting-specific mismatch between representations. &
Findings remain scoped to the target setting, such as VMI, a web application,
or a protocol stack. \\

\addlinespace[2pt]

Zero Trust Architecture~\cite{rose2020nist} &
Should this access request be allowed under current runtime trust signals? &
Per-request policy decision at an enforcement point. &
Decisions are made per request. They are not traced as design-level propagation
across trust boundaries. \\

\addlinespace[2pt]
\midrule

\textbf{TBSAM (this paper)} &
After syntactic validation succeeds at a trust boundary, which receiving-domain
security requirement remains unestablished? &
Unestablished requirement classified in an MDTBSG dimension
(Sec.~\ref{subsec:MDTBSG}). &
P1/P3 markers connect a gap at its preceding boundary to its appearance at
subsequent boundaries (Sec.~\ref{subsubsec:propagation}). Propagated gaps are
closed at their source. \\

\bottomrule
\end{tabularx}
\smallskip
\footnotesize
\textit{Note.} Input artifacts for each approach: STRIDE and TBSAM take
data-flow diagrams and architecture documents; SLSA takes build and
distribution pipeline specifications; semantic-gap analyses take
setting-specific artifacts (hypervisor state, application source,
protocol traces); ZTA takes runtime access requests and trust signals.
\end{table*}

The stages above analyze the SolarWinds attack one boundary at a time. This
subsection adds the cross-boundary view: how gaps at different boundaries
depend on one another and how they propagate along the exploit path.

\subsubsection{TBSG Propagation}
\label{subsubsec:propagation}

Figure~\ref{fig:propagation} summarizes the cross-boundary view of the
SolarWinds exploit path. The upper lane shows the artifact flow across
domains; the lower lanes align the same boundaries with the four MDTBSG
dimensions.

TB1's P1 Identity gap is the root. Once the attacker enters the build
environment as a legitimate user, later artifacts carry that unverified
assertion forward without creating new local Identity gaps at each
subsequent boundary.

The main propagation path appears between TB2 and TB3. At TB2, the build
process produces the signed artifact, so this is the first boundary where
the system can check whether the generated binary matches the intended
source-level behavior. No preceding boundary can establish that property
because the binary does not yet exist. The Interpretation gap therefore
originates at TB2 as a P1 gap. When the same signed artifact later reaches
customer systems, the Authenticode check at TB3 re-establishes the same
insufficient assertion over the same artifact. Placing the control at TB2
closes TB2's P1 gap and prevents the propagated TB3 gap; placing it only
at TB3 leaves the source of the gap---the build-output boundary---unaddressed.

The Temporal dimension follows a different pattern. It does not simply
reappear unchanged at each boundary; its window expands along the path.
At TB2, the gap lies between build-time checks and later use of the build
output. At TB3, the same timing problem extends across signing, distribution,
and customer installation. At TB4, SUNBURST's 12--14 day dormancy period
further separates earlier acceptance from later malicious behavior. The Spatial gap at TB4 is local rather than propagated. Even if TB1--TB3
were fully mitigated, those preceding boundaries would still not establish
whether an outbound destination is legitimate for Orion communication. The
outbound communication boundary must check that property itself, so TBSAM
treats the TB4 Spatial gap as an independent P1.

In TBSG terms, the SolarWinds attack is not four independent gaps at four
boundaries. It is a chain structure: an Identity root at TB1, an
Interpretation gap that originates at TB2 and propagates to TB3, a Temporal
gap that widens from TB2 to TB4, and an independent Spatial gap at TB4.
TBSAM keeps these cases separate so that propagated gaps are closed at their
preceding source and local gaps receive controls where they first appear.

\subsubsection{Comparison with STRIDE}
\label{subsubsec:stride_comparison}

STRIDE-based threat modeling and TBSAM use the same design artifacts but ask
different questions. STRIDE helps security architects identify threat
categories for each system element and map them to controls. TBSAM asks which
security property remains unestablished after those controls succeed, and how
that gap appears across trust boundaries. Table~\ref{tab:positioning}
summarizes this difference: for STRIDE, a selected control addresses the
finding at the element or boundary under review; for TBSAM, the successful
syntactic pass is the starting point, and P1/P3 markers link gaps across
preceding and subsequent boundaries. The remaining approaches in the
table---SLSA, semantic-gap analyses, and Zero Trust Architecture---are
discussed in Section~\ref{subsec:security_by_design}.

The SolarWinds case study makes this distinction concrete. A STRIDE-based
threat modeling pass over the boundary record identifies Tampering at TB2 and
TB3 and can lead to integrity controls such as code signing. This output is
correct: code signing addresses tampering. TBSAM records an additional
relation: TB3 repeats the same insufficient assertion already present at TB2.
A security architect can therefore see that a control placed at TB2 can also
close the propagated gap at TB3. TBSAM also separates the locally originating
Spatial gap at TB4 from the inherited Interpretation gap at TB3. These P1,
P2, and P3 markers help place controls where gaps originate, not only where
they become visible.

The two approaches therefore complement each other. STRIDE names the threat
category. TBSAM names the unestablished security property behind the accepted
artifact and traces whether the gap starts locally or comes from a preceding
boundary.

\section{Discussion}
\label{sec:discussion}

This section discusses the implications and limitations of TBSAM.
Section~\ref{subsec:security_by_design} explains how TBSAM supports
Security-by-Design and relates to STRIDE, Zero Trust Architecture (ZTA), and
SLSA. Section~\ref{subsec:limitations} describes the limitations of this work
and directions for future research.

\subsection{Toward Security-by-Design}
\label{subsec:security_by_design}

A defense-in-depth process treats syntactic validation as the first line of
defense. The residual question---which security property the receiving domain
still cannot assume after that line succeeds---is what TBSAM makes explicit
before implementation. Instead of adding another parser, filter, or check at
the boundary, TBSAM records what the selected check leaves open. As
Sections~\ref{subsec:MDTBSG} and~\ref{subsec:overview_TBSAM} show, MDTBSG
frames this question around the receiving domain's security properties, while
TBSAM starts from the controls selected during threat modeling and examines
what they leave unestablished at the boundary.

Web applications, software supply chains, cloud and identity infrastructure,
and operating systems share the same boundary-level structure: a successful
check does not establish the security meaning the receiving domain requires.
The 75 incidents analyzed in Section~\ref{subsec:incident_analysis} span these
areas, which suggests that TBSG is a recurring design-level problem rather
than a set of isolated bugs. TBSAM gives security architects a structured way
to expose this problem before implementation and map it to architectural
controls.

ZTA removes implicit trust at runtime enforcement points; TBSAM surfaces
boundary-crossing assumptions during system design, before they are encoded as
enforcement decisions~\cite{rose2020nist}. A credential accepted at one
boundary may later influence artifacts or requests evaluated at subsequent
boundaries. TBSAM's P3 markers record such chains so that a policy designer
can see which assumptions a runtime policy may need to re-evaluate. TBSAM does
not define ZTA policy, but it makes these boundary-crossing assumptions
explicit before implementation.

Supply-chain frameworks such as SLSA~\cite{slsa} prescribe integrity levels
for build provenance and, at higher levels, can require controls such as
source-to-binary attestation. This aligns with the control TBSAM selects for
TB2 in the SolarWinds reconstruction, where the boundary record shows that the
build-output boundary leaves source-to-binary behavior unestablished. SLSA
specifies which controls a pipeline should deploy to reach a target level;
TBSAM identifies where the residual gap first appears in the boundary record
and traces how it propagates when that gap remains open.

\subsection{Limitations and Future Work}
\label{subsec:limitations}

This work has several limitations.

TBSAM requires the security architect to identify trust boundaries, determine
the security properties required by the receiving domain, and assign P1, P2,
and P3 priorities. These steps depend on the quality of the design documents
and on domain knowledge. Different security architects may therefore identify
different gaps or assign different priorities for the same boundary. The
dimension-specific questions in this paper reduce this ambiguity, but they do
not remove it. Future work should provide clearer decision rules, for example
by using attack trees, abuse cases, or explicit threat scenarios as additional
input.

The incident set used in this paper is based on public incident reports, CVE
descriptions, and technical analyses. Incidents that were not publicly
disclosed, or for which technical details are limited, are not included. This
may affect how often each MDTBSG dimension appears in our analysis, because
public reports may make some gaps easier to observe than others. In addition,
MDTBSG currently consists of four dimensions: Identity, Spatial, Temporal, and
Interpretation. These dimensions capture the recurring patterns found in the
incident set, but they should not be treated as a closed list. Future work
should extend the incident set and examine whether additional dimensions
appear in other systems or less visible incidents.

The case studies in this paper are retrospective analyses. They show which
trust boundaries, semantic gaps, and architectural controls TBSAM could have
exposed during design, but they do not measure its effect in a live design
process. Future work should apply TBSAM to systems under active development
and measure whether it helps security architects find trust-boundary flaws
earlier, choose better architectural controls, and reduce exploitable attack
paths before implementation.

TBSAM identifies semantic gaps and maps them to architectural controls, but it
does not define how those controls should be enforced or monitored at runtime.
Future work should translate TBSAM outputs into runtime policies, monitoring
rules, and enforcement points. Another direction is partial tool support for
Stages~1 and~2 when system designs are machine-readable, for example by
extracting trust boundaries from data-flow diagrams, API gateway
configurations, or service-mesh policies, then presenting candidate boundaries
and dimensions to the security architect for priority assignment.

\section{Conclusion}
\label{sec:conclusion}

This paper defined Trust Boundary Semantic Gap (TBSG), a
design-level condition in which an artifact passes correctly implemented
syntactic validation at a trust boundary, but the assertions established
by that pass remain insufficient for the receiving domain. TBSG shifts
the analysis point from absent or broken validation to what remains
unestablished after a syntactic pass succeeds.

Through an analysis of publicly reported security incidents, we
organized recurring forms of semantic misalignment into four dimensions:
Identity, Spatial, Temporal, and Interpretation. We then developed Trust
Boundary Semantic Analysis and Mitigation (TBSAM), a design-time
framework that identifies TBSGs from design specifications, prioritizes
the resulting gaps, and maps them to candidate architectural controls.
The SolarWinds/SUNBURST case study showed how TBSAM can make
receiving-domain assumptions explicit along a real exploit path and
complement STRIDE by preserving cross-boundary propagation.

Systems may correctly check signatures, tokens, schemas, protocols, and formats while still relying on semantic assumptions that those mechanisms do not establish. Making these assumptions explicit during system design is a practical step toward Security-by-Design.

\bibliographystyle{IEEEtran} 
\bibliography{citation}      

\appendices
\suppressfloats[t]
\section{Incident Analysis}

Table~\ref{tab:Analysis} maps the 75 incidents used in this paper to the TBSG dimensions implicated at the exploited trust boundary. Each label follows the coding rule described in Section~\ref{subsec:incident_analysis}: the public record must support that an artifact crossed a trust boundary, that the receiving domain accepted the artifact after syntactic validation passed, and that the exploit depended on a receiving-domain security property not established by that pass. The cited source for each incident is the most specific public record available for the boundary-level classification, such as a CVE record, government advisory, peer-reviewed paper, vendor security advisory, incident report, or threat-intelligence analysis. When no CVE or government advisory is available, we use the primary public incident report or vendor analysis that documents the exploited path. Empty cells indicate dimensions for which the three-fact criterion was not met in the public record; they are part of the coding result, not an absence of analysis.

\providecommand{\YES}{\ensuremath{\bullet}}
\providecommand{\NO}{\textcolor{gray}{---}}

\begin{table*}[t]
\centering
\caption{MDTBSG Dimension Mapping for the 75-Incident Analysis (\YES{}\,=\,implicated; \NO{}\,=\,not implicated.)}
\label{tab:Analysis}
\scriptsize
\setlength{\tabcolsep}{1.8pt}
\renewcommand{\arraystretch}{0.95}

\begin{minipage}[t]{0.49\textwidth}
\begin{tabular}{@{}>{\raggedleft\arraybackslash}p{3.2mm}
                  >{\raggedright\arraybackslash}p{58mm}
                  cccc@{}}
\toprule
\textbf{\#} & \textbf{Incident} & \textbf{Id} & \textbf{Sp} & \textbf{Te} & \textbf{In} \\
\midrule
1  & SolarWinds / SUNBURST~\cite{crowdstrike2024rca} & \YES & \YES & \YES & \YES \\

2  & Kaseya VSA & \YES & \YES & \NO  & \YES \\

3  & NotPetya / M.E.Doc & \YES & \YES & \YES & \YES \\

4  & XZ Utils Backdoor& \YES & \YES & \YES & \YES \\

5  & Polyfill.io & \YES & \YES & \NO  & \YES \\

6  & 3CX Desktop App & \YES & \YES & \YES & \YES \\

7  & Ledger Connect Kit & \YES & \YES & \NO  & \YES \\

8  & BadBox Android & \YES & \YES & \NO  & \YES \\

9  & MagicWeb ADFS DLL & \YES & \YES & \YES & \YES \\

10 & Codecov Bash Uploader & \YES & \YES & \YES & \YES \\

11 & tj-actions/changed-files & \YES & \YES & \YES & \YES \\

12 & Shai-Hulud npm Worm & \YES & \YES & \YES & \YES \\

13 & Cyberhaven Chrome Ext. & \YES & \YES & \YES & \YES \\

\midrule
14 & Okta / LAPSUS\$ via Sitel & \YES & \YES & \YES & \NO  \\

15 & GitHub OAuth Token Theft & \YES & \YES & \YES & \NO  \\

16 & Snowflake Customer Instances & \YES & \YES & \YES & \NO  \\

17 & Microsoft Midnight Blizzard~\cite{msrc2024midnight} & \YES & \YES & \YES & \NO  \\

18 & Colonial Pipeline~\cite{cisa2021colonial} & \YES & \YES & \YES & \NO  \\

19 & Nvidia LAPSUS\$ & \YES & \YES & \YES & \NO  \\

20 & Storm-0558 Exchange Online~\cite{msrc2023storm0558,csrb2024storm0558} & \YES & \YES & \YES & \NO  \\

21 & Entra ID Actor Token~\cite{cve202555241} & \YES & \YES & \YES & \NO  \\

22 & Okta Support System HAR Leak & \YES & \YES & \YES & \YES \\

\midrule
23 & MS Exchange ProxyLogon~\cite{nvd_cve202126855} & \YES & \YES & \NO  & \YES \\

24 & Confluence Widget Connector~\cite{nvd_cve20193396} & \YES & \YES & \NO  & \YES \\

25 & Spring4Shell~\cite{nvd_cve202222965} & \YES & \NO  & \NO  & \YES \\

26 & Log4Shell~\cite{nvd_cve202144228} & \YES & \YES & \NO  & \YES \\

27 & Capital One SSRF~\cite{doj2019capitalone_arrest, 10.1145/3546068} & \YES & \YES & \YES & \YES \\

28 & Equifax Struts2 RCE~\cite{us2018equifax} & \YES & \NO  & \NO  & \YES \\

\midrule
29 & Docker Symlink TOCTOU & \NO  & \YES & \YES & \NO  \\

30 & Tomcat JSP TOCTOU~\cite{cve2017tomcatjsp} & \NO  & \YES & \NO  & \YES \\

31 & Tesla Model~3 Gateway & \YES & \YES & \YES & \NO  \\

32 & WordPress DNS Rebinding & \YES & \YES & \YES & \NO  \\

33 & AWS DynamoDB DNS Race & \NO  & \YES & \YES & \YES \\

34 & Requests Auth Header Leak~\cite{cve201818074} & \NO  & \YES & \NO  & \YES \\

35 & Quick Heal AV Symlink~\cite{cve202231466} & \NO  & \YES & \YES & \NO  \\

\midrule
36 & Node.js DLL Hijacking~\cite{cve202232223} & \NO  & \YES & \NO  & \YES \\

37 & Windows RDP DLL Hijacking~\cite{msrc2024cve202438077} & \YES & \NO  & \NO  & \YES \\

38 & ToneShell DLL Side-loading~\cite{unit422023toneshell} & \YES & \NO  & \NO  & \YES \\

\bottomrule
\end{tabular}
\end{minipage}\hfill
\begin{minipage}[t]{0.49\textwidth}
\begin{tabular}{@{}>{\raggedleft\arraybackslash}p{3.2mm}
                  >{\raggedright\arraybackslash}p{58mm}
                  cccc@{}}
\toprule
\textbf{\#} & \textbf{Incident} & \textbf{Id} & \textbf{Sp} & \textbf{Te} & \textbf{In} \\
\midrule
39 & Salesloft Drift Credential Leak & \YES & \YES & \YES & \NO  \\

40 & AWS AppSync Confused Deputy~\cite{frichette2022appsync} & \YES & \YES & \NO  & \YES \\

\midrule
41 & Android StrandHogg 2.0~\cite{nvd_cve20200096} & \YES & \YES & \NO  & \YES \\

42 & ByBit Safe\{Wallet\} & \YES & \YES & \NO  & \YES \\

43 & WalletConnect UI Spoofing~\cite{walletconnect2023spoofing} & \NO  & \NO  & \NO  & \YES \\

44 & Trojan Source Bidi~\cite{bsw2021trojansource} & \NO  & \NO  & \NO  & \YES \\

\midrule
45 & Android Camera Intent & \YES & \YES & \NO  & \NO  \\

46 & Android Bundle Feng Shui~\cite{nvd_cve20210928} & \YES & \YES & \NO  & \YES \\

47 & WorkSource / Pinduoduo~\cite{nvd_cve202320963} & \YES & \YES & \NO  & \YES \\

48 & Dirty Stream & \YES & \YES & \NO  & \YES \\

49 & Migraine macOS SIP & \YES & \YES & \YES & \YES \\

50 & FORCEDENTRY iMessage~\cite{gpz2021forcedentry} & \YES & \YES & \NO  & \YES \\

\midrule
51 & CircleCI SSO Token & \YES & \YES & \YES & \NO  \\

\midrule
52 & TeamCity Auth Bypass~\cite{nvd_cve202427198} & \YES & \YES & \YES & \YES \\

53 & ScreenConnect Setup~\cite{nvd_cve20241709} & \YES & \YES & \YES & \YES \\

54 & Confluence Admin~\cite{nvd_cve202322515} & \YES & \YES & \NO  & \YES \\

55 & PaperCut Auth Bypass~\cite{nvd_cve202327350} & \YES & \YES & \YES & \YES \\

\midrule
56 & MOVEit Transfer SQLi~\cite{nvd_cve202334362} & \YES & \YES & \YES & \YES \\

57 & Confluence OGNL~\cite{nvd_cve202226134} & \YES & \YES & \YES & \YES \\

58 & GoAnywhere MFT~\cite{nvd_cve20230669} & \YES & \YES & \YES & \YES \\

59 & Struts2 XStream~\cite{apache2017struts9805} & \YES & \NO  & \NO  & \YES \\

60 & Bitbucket CMDi~\cite{nvd_cve202236804} & \NO  & \NO  & \NO  & \YES \\

61 & GitLab ExifTool~\cite{gitlab2021exiftool} & \NO  & \NO  & \NO  & \YES \\

\midrule
62 & Ivanti Connect Secure~\cite{nvd_cve202346805} & \YES & \NO  & \NO  & \YES \\

63 & FortiOS SSL VPN~\cite{nvd_cve202242475} & \YES & \NO  & \NO  & \YES \\

64 & Barracuda ESG TAR~\cite{nvd_cve20232868} & \YES & \YES & \NO  & \YES \\

65 & Cisco IOS XE Web UI & \NO  & \NO  & \NO  & \YES \\

\midrule
66 & AWS Lambda Region Evasion & \YES & \YES & \YES & \YES \\

67 & Azure AD Legacy Auth & \YES & \YES & \NO  & \YES \\

68 & GCP Cloud Shell OAuth & \YES & \YES & \YES & \NO  \\

\midrule
69 & CrowdStrike CF291~\cite{crowdstrike2024rca} & \YES & \YES & \YES & \YES \\

70 & Confluence JMX Deserialization RCE.~\cite{nvd_cve202322508} & \YES & \YES & \YES & \YES \\

71 & Vault Namespace~\cite{nvd_cve20230620} & \YES & \YES & \YES & \YES \\

72 & Jenkins CLI File Read~\cite{sonatype2024jenkins} & \YES & \YES & \YES & \YES \\

\midrule
73 & PyTorch torchtriton Confusion & \YES & \NO  & \NO  & \YES \\

74 & Ultralytics PyPI Poisoning & \YES & \YES & \NO  & \YES \\

75 & Nx s1ngularity Pwn Request & \YES & \YES & \YES & \YES \\

\midrule
\multicolumn{2}{@{}l}{\textbf{Total (75)}}         & \textbf{64} & \textbf{62} & \textbf{40} & \textbf{59} \\
\multicolumn{2}{@{}l}{\textbf{Prev.\ (\%)}}         & \textbf{85.3} & \textbf{82.7} & \textbf{53.3} & \textbf{78.7} \\
\bottomrule
\end{tabular}
\end{minipage}
\end{table*}

\section{TBSAM Boundary Record: SolarWinds/SUNBURST}
\label{app:solarwinds_boundary_record}

This appendix presents the Stage~1 and Stage~2 outputs of TBSAM applied to the SolarWinds Orion software supply chain. Because SolarWinds has not published a design-level architecture document for its build and distribution pipeline, we reconstruct the trust domains, trust boundaries, and syntactic validation mechanisms from post-incident analyses published by CrowdStrike~\cite{crowdstrike_sunspot}, Mandiant~\cite{mandiant_sunburst}, ReversingLabs~\cite{reversinglabs_sunburst}, and CISA~\cite{cisa_aa20352a}. Table~\ref{tab:sw_trust_domains} lists the reconstructed trust domains. Table~\ref{tab:sw_boundary_record} presents the boundary record constructed in Stage~1, with candidate dimensions from Stage~2 marked in the rightmost four columns.

\begin{table*}[!t]
\centering
\caption{Reconstructed Trust Domains in SolarWinds Orion Supply Chain}
\label{tab:sw_trust_domains}
\renewcommand{\arraystretch}{1.2}
\footnotesize
\begin{tabularx}{\textwidth}{clX}
\toprule
\textbf{Domain} & \textbf{Component} & \textbf{Security Assumption} \\
\midrule
$D_{\text{ext}}$   & External network      & Untrusted; includes the attacker's initial access vector \\
$D_{\text{build}}$ & Build environment     & Trusted compilation environment; source code integrity assumed; build output is signed before release \\
$D_{\text{dist}}$  & Distribution server   & Hosts signed update packages; customers retrieve packages through the official update channel \\
$D_{\text{cust}}$  & Customer system       & Operates the deployed Orion instance; trusts packages signed by the SolarWinds certificate \\
$D_{\text{c2}}$    & C2 infrastructure     & Attacker-controlled; receives beacons from compromised customer systems via DNS and HTTPS \\
\bottomrule
\end{tabularx}
\end{table*}

\begin{table*}[t]
\centering
\caption{Boundary Record and Candidate Dimensions for SolarWinds/SUNBURST Supply Chain}
\label{tab:sw_boundary_record}
\renewcommand{\arraystretch}{1.25}
\footnotesize
\begin{tabularx}{\textwidth}{cllXcccc}
\toprule
\textbf{TB} & \textbf{Crossing} & \textbf{Artifact} & \textbf{Syntactic Validation ($\Phi$)} & \textbf{Id} & \textbf{Sp} & \textbf{Te} & \textbf{In} \\
\midrule

\multirow{3}{*}{TB1} & \multirow{3}{*}{$D_{\text{ext}} \!\to\! D_{\text{build}}$} & \multirow{3}{*}{\shortstack[l]{Remote access\\session\\(credential)}} 
  & Credential authentication (password / MFA) & $\bullet$ & & & \\
\cmidrule(lr){4-8}
  & & & VPN / remote access gateway session validation & $\bullet$ & & & \\
\cmidrule(lr){4-8}
  & & & Network perimeter access control (firewall rules) & & & & \\
\midrule

\multirow{4}{*}{TB2} & \multirow{4}{*}{$D_{\text{build}} \!\to\! D_{\text{dist}}$} & \multirow{4}{*}{\shortstack[l]{Signed build\\artifact\\(.dll / .msp)}} 
  & Code-signing signature generation (Authenticode) & & & & $\bullet$ \\
\cmidrule(lr){4-8}
  & & & Build compilation success check (MSBuild exit code) & & & & $\bullet$ \\
\cmidrule(lr){4-8}
  & & & Build output hash verification (MD5 / SHA-256) & & & $\bullet$ & \\
\cmidrule(lr){4-8}
  & & & Package format and metadata validation & & & & \\
\midrule

\multirow{3}{*}{TB3} & \multirow{3}{*}{$D_{\text{dist}} \!\to\! D_{\text{cust}}$} & \multirow{3}{*}{\shortstack[l]{Signed update\\package\\(.msp)}} 
  & Code-signing certificate and signature verification (Authenticode) & & & & $\bullet$ \\
\cmidrule(lr){4-8}
  & & & Update channel validation (official SolarWinds download URL) & & $\bullet$ & & \\
\cmidrule(lr){4-8}
  & & & Package integrity check (hash comparison) & & & $\bullet$ & \\
\midrule

\multirow{3}{*}{TB4} & \multirow{3}{*}{$D_{\text{cust}} \!\to\! D_{\text{c2}}$} & \multirow{3}{*}{\shortstack[l]{DNS query +\\HTTPS beacon\\(HTTP GET/PUT)}} 
  & DNS resolution and response validation & & $\bullet$ & & \\
\cmidrule(lr){4-8}
  & & & HTTPS/TLS certificate verification & & $\bullet$ & & \\
\cmidrule(lr){4-8}
  & & & HTTP protocol compliance (method, header, body format) & & & $\bullet$ & \\
\bottomrule
\end{tabularx}

\smallskip\noindent\footnotesize
\textbf{Id} = Identity, \textbf{Sp} = Spatial, \textbf{Te} = Temporal, \textbf{In} = Interpretation.\\
$\bullet$ indicates that the assertion established by $\Phi(A)$ through this validation item does not by itself satisfy $R_j(A)$ in the corresponding dimension. An empty cell indicates that the validation item does not independently produce a candidate gap in that dimension.\\
\textsl{Source Attribution:} TB1 access mechanisms are reconstructed from CISA~AA20-352A~\cite{cisa_aa20352a} and CrowdStrike's SUNSPOT analysis~\cite{crowdstrike_sunspot}; TB2 signing and build-integrity mechanisms from ReversingLabs~\cite{reversinglabs_sunburst} and CrowdStrike~\cite{crowdstrike_sunspot}; TB3 customer-side verification from the SolarWinds Security Advisory and Mandiant~\cite{mandiant_sunburst}; and TB4 C2 protocols from Mandiant's SUNBURST technical analysis~\cite{mandiant_sunburst}.
\end{table*}
\end{document}